\documentstyle[12pt]{article}
\newcommand{\refs}[1]{(\ref{#1})}
\def\mtx#1{\quad\hbox{{#1}}\quad}
\def\es{\!=\!}

\def\pa{\partial}

\def\ov{\over}
\def\ha{{1\over 2}}

\def\sg{\sqrt{g}}

\def\>{\rangle}
\def\<{\langle}
\def\es{\!=\!}

\def\La{\Lambda}
\def\la{\lambda}
\def\be{\beta}

\def\si{\sigma}
\def\al{\alpha}
\def\de{\delta}
\def\om{\omega}
\def\Om{\Omega}

\def\ar{\rightarrow}

\def\bib{\bibitem}
\def\D{{\cal D}}

\def\pan{\par\noindent}

\def\beq{\begin{equation}}
\def\eeq{\end{equation}}
\def\bed{\begin{displaymath}}
\def\eed{\end{displaymath}}
\def\beqq{\begin{eqnarray}}
\def\eeqq{\end{eqnarray}}
\def\bedd{\begin{eqnarray*}}
\def\eedd{\end{eqnarray*}}

\advance\hoffset by -.35in
\begin{document}

\begin{flushright}
%hep-th XXXXXXX\\
{\bf DIAS-STP 96-11\\December 1996}
\end{flushright}

\def\sqr#1#2{{\vcenter{\vbox{\hrule height.#2pt
\hbox{\vrule width.#2pt height#1pt \kern#1pt
\vrule width.#2pt}
\hrule height.#2pt}}}}
\def\square{\mathchoice\sqr34\sqr34\sqr{2.1}3\sqr{1.5}3}

\begin{center}\huge{\bf{Weyl-Gauging and Conformal Invariance}}
\end{center}
\vspace{1cm}

\begin{center}
A. Iorio, L. O'Raifeartaigh,
I. Sachs and C. Wiesendanger \\
{\it Dublin Institute for Advanced Studies\\
School of Theoretical Physics\\
10 Burlington Road, Dublin 4, Ireland}
\end{center}
\vspace{1cm}

\begin{abstract}
Scale-invariant actions in arbitrary dimensions 
are investigated in curved space to
clarify the relation between scale-, Weyl- and conformal
invariance on the classical level. The global Weyl-group is gauged. Then
the class of actions is determined for which
Weyl-gauging may be replaced by
a suitable coupling to the curvature (Ricci gauging).
It is shown that this class is exactly the class of actions 
which are conformally invariant
in flat space. The procedure yields a simple algebraic criterion
for conformal invariance and produces the improved energy-momentum
tensor in conformally invariant theories in a systematic way. 
It also provides a simple and fundamental 
connection between Weyl-anomalies and central extensions in 
two dimensions. In particular, the subset of scale-invariant 
Lagrangians for fields of 
arbitrary spin, in any dimension, which are conformally invariant 
is given. An example of a quadratic action 
for which scale-invariance
does not imply conformal invariance is constructed.
\end{abstract}

\clearpage

\section{Introduction}

It has been known for a long time that
the symmetric (Belinfante) energy-momentum tensor may be obtained
by writing the flat-space action $A$ in a diffeomorphic-invariant
manner ${\cal A}$ and by computing its derivative with
respect to the metric tensor \cite{Ros}
\beq
T_{\mu\nu}={4\pi\over\sqrt{g}}{\delta {\cal A}\over\delta g^{\mu\nu}}.
\label{1.1}
\eeq
For scale-invariant theories the trace $T^\mu_{\;\;\mu}$
of the symmetric
energy-momentum tensor should be zero in the flat-space limit
but in practice this is not always
the case and the symmetric $T_{\mu\nu}$ has
to be replaced by a so-called 'improved' energy-momentum tensor 
\cite{cacoja}.
In general, the improvement is found in a rather ad hoc manner but
it has been observed that for some scale-invariant scalar 
theories the improved energy-momentum tensor may also be obtained using
(\ref{1.1}) provided that a suitable extra coupling of the fields
to the Ricci scalar $R$ is introduced \cite{cacoja,Jade}. 
Furthermore, in certain 
two-dimensional conformal field theories this procedure
allows the Virasoro centre $c$ to be computed in a simple
manner using the identity \cite{Aff}
\beq
\<T_{\;\;\mu}^\mu\>=-{c\over 12}R. \label{1.2}
\eeq
To our knowledge it has however not been established so far why, 
and under what conditions this procedure works, and what form the coupling to $R^\mu\,_{\nu\rho\si}$ should take for fields of arbitrary spin. The purpose of
the present paper is to fill this gap.

Our strategy is as follows: First we 
convert the rigid scale-invariance to rigid Weyl-invariance and
then promote the latter to local Weyl-invariance by gauging in
the standard manner. This permits the question to be
formulated more precisely as to how and when
the Weyl gauge-potential $W_\mu$ may be replaced by curvature-terms.
We find that the necessary and sufficient
condition for such a replacement is that the theory should
be conformally 
invariant and that the general coupling is to
the Ricci tensor $R_{\mu\nu}$ rather than the Ricci scalar $R$.

Hence, we obtain a new criterion for conformal invariance. In the
infinitesimal limit it reduces to the 
criterion for conformal invariance which was
given in \cite{coja,P}, namely
that the so-called virial current $j_{\mu}$ be the divergence of
a tensor, $j_{\mu}=\pa_\nu J_\mu\,^{\nu}$. However, we can go further 
considering finite 
local Weyl-transformations. This allows us to identify
the tensor $J_\mu\,^{\nu}$ and thus to reduce the condition for conformal
invariance to an {\it algebraic condition} which permits a systematic  
analysis. We then obtain the constraints for conformal invariance and the 
coupling to the Ricci-tensor for Lagrangians for arbitrary spin in any 
dimensions. For spin zero we also exhibit an action which is
scale-invariant and quadratic in the derivatives of the fields but
is not conformally invariant.

The paper is organized as follows. In section $2$ we illustrate the
coupling to the Ricci scalar by means of
the simplest non-trivial conformally invariant theory,
namely the Liouville theory, in both the classical and
quantum cases. In section $3$ we establish the equivalence of rigid
scale-invariance and rigid Weyl-invariance in the
diffeomorphic context. The latter is then gauged in the standard
manner by 
introducing a Weyl-potential $W_\mu$. In sections $4$ and $5$ we
analyse under what conditions $W_\mu$ 
may be replaced by a particular combination of the Ricci 
tensor $R_{\mu\nu}$ and the Ricci scalar $R$ and show 
that these conditions are equivalent to conformal invariance in 
flat space. The relation between our results and
those of \cite{coja,P} is also discussed there. In section $6$ we discuss 
in detail the role of central terms in the conformal variation of the 
Lagrangian and relate them to gravitational 
anomalies. The extension of our approach to quantum systems is also 
explained in that section. Finally, in section $7$ we find the set of 
scale-invariant Lagrangians which are conformally invariant for fields of 
arbitrary spin and construct the abovementioned counter example. 
Section $8$ contains the conclusions. 
Our conventions are those of Birrell and Davies 
\cite{bide}.

\section{A Two-dimensional Illustration}

We first illustrate the use of non-minimal coupling to the curvature 
(Ricci gauging) by
means of the two-dimensional Liouville theory with action
\beq
{\cal A}=\frac{1}{4\pi}\int \sqrt{g}\,\Bigl({1 \over 2}\pa^\mu \phi\;
\pa_\mu \phi-m^2e^\phi+R\phi\Bigr),  \label{2.1}
\eeq
where we have discarded a purely geometrical Polyakov term
which means that 
the action \refs{2.1} will be Weyl-invariant only up
to field-independent 
terms. Varying \refs{2.1} as in \refs{1.1} one then obtains
\beq
T_{\mu\nu}=
{1\over 2}\pa_{\mu}\phi\;\pa_{\nu}\phi-{1 \over 2}g_{\mu\nu}
\left(\ha\pa^\rho \phi\;\pa_\rho \phi-m^2e^\phi\right)
+\left(g_{\mu\nu}\nabla^2-
\nabla_{\mu}\pa_{\nu}\right)\phi.  \label{2.2}
\eeq
Note that 
$\left(g_{\mu\nu}\nabla^2-\nabla_{\mu}\pa_{\nu}\right)\phi$ is
the well-known improvement term \cite{cacoja}. It is clear that
it originates from the $R\phi$-term in the action. This term
survives in the flat space limit even though $R\phi$ itself
vanishes. Hence, the processes of varying with respect
to $g^{\mu\nu}$ and taking the flat limit do not commute. 
The trace of the energy-momentum tensor \refs{2.2} is then
\beq
T^\mu_{\;\;\mu}=m^2e^\phi +\nabla^2 \phi , \label{2.3}
\eeq
which reduces to $R$ on the mass-shell and vanishes in the
flat limit.

In the quantum case the situation is a little more subtle.
Because of renormalization the classical 
coefficient of the coupling-term $R\phi$ must be modified. That is,
we write 
$\al R\phi$ with $\al$ a
constant to be fixed according to the trace condition \refs{1.2}. 
For arbitrary $\al$ the variation of $Z[g]=\int \D\phi\; 
e^{\frac{i}{\hbar}{\cal A}}$ 
then leads to the exact result
\beq
\<T^\mu_{\;\;\mu}\>
={\sqrt{g}\ov Z[g]}{\de Z[g]\over \de\sqrt{g}}
=-{1\ov 12}\left(\hbar R +\al^2 R
+12(\al-1-{\hbar \over 2})\nabla^2 \<\phi\>\right),  \label{2.5}
\eeq
where the first term is due to the scale-anomaly in the measure. 
Since scale-invariance requires 
that
$\<T^\mu_{\;\;\mu}\>$
vanish in the flat limit the constant $\al$ must be
$\al=1+{\hbar \over 2}$. Then the right-hand side of
(\ref{2.5}) reduces to $cR$ where $c=\hbar +12(1+{\hbar \over 2})^2$,
which is just the expression for the Virasoro centre 
obtained earlier by canonical methods \cite{cuth}. We note in passing that variations of the model described above have been discussed extensively in connection with dilaton gravity and string theory \cite{Callan,Polyakov}.

\section{Rigid Scale- and Rigid Weyl-Invariance} 
In this section we convert rigid scale-invariance into rigid 
Weyl-invariance which we then promote to a local symmetry by gauging 
in the standard manner. For this we first consider the 
diffeomorphic version of Poincar\'e-invariant
flat-space theories. For simplicity we shall consider actions
$A$ for which the Lagrangian contains only the fields, denoted
generically by $\phi$, and their first derivatives
\beq
A=\int d^n x L(\phi,\pa_a\phi).  \label{3.1}
\eeq
We make $A$ diffeomorphic
invariant by letting
\beq
A\rightarrow {\cal A}=
\int d^n x\sqrt{g}\;L(\phi,e^{\mu}_a\nabla_{\mu}\phi),  \label{3.2}
\eeq
where  $\nabla_{\mu}$ is the diffeomorphism-covariant derivative,
$e^{\mu}_a$ is the Vielbein defined by
$e_{\mu a}\,e^a_{\nu}=g_{\mu\nu}$ and $g_{\mu\nu}$ is the metric.
More explicitly, for a field  $\phi^m$ with spin indices $m$
the covariant derivative may be written as
\beq
\bigl(\nabla_{\mu}\phi\bigr)^m
=\pa_{\mu}\phi^m +S_{\mu,n}^m\phi^n,  \label{3.3}
\eeq
where the spin connection $S_\mu$ is given by
\beq
S_{\mu,n}^m
=-S_{\mu}^{ab} \Bigl(\Sigma_{ab}\Bigr)^m_{\;\;n},\qquad
S_{\mu}^{ab}= e^a_\la
\bigl(g^\la\,_\tau\pa_\mu
+\Gamma_{\mu\tau}^\la\bigr) e^{\tau a}  \label{3.4}
\eeq
and the $\Sigma_{ab}$ are the generators of the
Lorentz transformations.
\vskip 0.3truecm \noindent
{\it Rigid Scale-Invariance
vs. Rigid Weyl-Invariance} :\par\noindent
By a rigid scale-transformation in flat space we mean
a transformation of the Cartesian coordinates and of the fields
\beq
x^{a}\ar  e^{\om} x^{a} \qquad \phi\ar 
e^{d_\phi\om} \phi,  \label{3.5}
\eeq
where $\om$ is a real constant and $d_\phi$ is the dimension
of the field. In practice, the dimension
of the fields are canonical, ie. determined by
the kinetic part of the Lagrangian.
Now, because of diffeomorphism invariance the transformations 
(\ref{3.5}) are equivalent to the transformations
\beq
e_{\mu}^m\ar  e^{\om} e_{\mu}^m
\quad \hbox{and} \quad \phi\ar e^{d_\phi\om}\phi
\label{3.6}
\eeq
{\it with no change of coordinates}. In other words, the scaling can
be transferred to the Vielbein and thence to the metric and to
$\sqrt{g}$, which scales as $e^{n\om}$. The transformations (\ref{3.6})
are called rigid Weyl-transformations and thus any action which is
invariant with respect to rigid scale-transformations in the
flat limit is rigid Weyl-invariant in curvilinear
coordinates. Rigid Weyl-invariance is therefore the natural way of
generalizing rigid scale-invariance to curvilinear coordinates.
\vskip 0.3truecm \noindent
{\it Weyl-gauging} :\par\noindent
To see under what conditions a theory can be made locally Weyl-invariant 
by a non-minimal coupling to curvature-terms we first consider 
standard Weyl-gauging. Because rigid Weyl-invariance 
is an inner symmetry it can always be 
converted to a local symmetry by gauging in the
usual manner. The purpose of Weyl-gauging is
to compensate for the inhomogeneous terms that arise in the
derivatives of the fields for local Weyl-transformations, 
$\phi(x)\ar  e^{d_\phi\om(x)}\phi(x)$.

Let us first consider scalar fields. In that case we have
\beq
\pa_\mu(e^{d_\phi\om(x)}\phi(x))
= e^{d_\phi\om(x)}
(\pa_\mu +d_\phi\,\pa_\mu\om(x))\phi(x)) 
\label{4.1}
\eeq
and hence Weyl-gauging consists of letting
\beq
\pa_\mu \phi \ar  D_\mu \phi, \quad \hbox{where} \quad
D_\mu=\pa_\mu+{d_\phi}W_\mu \quad \hbox{and}\quad
W_\mu\ar  W_\mu-\pa_\mu\om.  \label{4.2}
\eeq
The situation is quite analogous to that in electrodynamics and
indeed  historically Weyl-gauging preceded and motivated
electrodynamic gauging. 
However, it differs from electrodynamics in that there
is no imaginary factor $i$. An important consequence of this is
that the current associated with $W_\mu$ is not of the electromagnetic
form $i(\phi^*\pa^\mu \phi-\phi\,\pa^\mu \phi^*)$ but
$(\phi^*\pa^\mu \phi+\phi\,\pa^\mu \phi^*)=\pa^\mu(\phi^*\phi)$.
Thus, it is not zero for real fields and is a total derivative.

The situation for higher spin fields is a little
more complicated because  the ordinary derivative
$\pa_\mu$ is replaced by the diffeomorphic
covariant derivative $\nabla_\mu$
and the latter varies with respect to local Weyl-transformations. To
see how it varies we recall that in the Vielbein formalism it takes
the form $\nabla_\mu=\pa_\mu+S_\mu$ where
$S_\mu$ is given in (\ref{3.4}). Since the Vielbein $e_\tau^b$
is a covariant vector it has scale-dimension
$d_e=1$ and hence the Weyl-variation of $\nabla_\mu$ is
\beq
\Delta\nabla_\mu=-2\Sigma_\mu\,^\nu \pa_\nu\om.  \label{4.3}
\eeq 
Taking the variations of $\nabla_\mu$ and
of the field into account we see that for general
spin fields we must Weyl-gauge according to
\beq
\nabla_\mu\quad \ar  \quad
\nabla_\mu+\La_{\;\;\mu}^\nu W_\nu,  \label{4.4}
\eeq
where
\beq
\La_{\;\;\mu}^\nu={d_\phi}g_{\;\;\mu}^\nu+
2\Sigma_{\;\;\mu}^\nu \quad \hbox{and}\quad
W_\nu\ar  W_\nu-\pa_\nu\om.  \label{4.5}
\eeq
The operator $\La_{\nu\mu}$ is the same as that
entering in the virial current \cite{coja}. Its spin component 
$\Sigma_{\mu\nu}$ is due to the spin connection. In the next section 
we will see under what conditions the Weyl-gauging is equivalent to a 
non-minimal coupling to curvature.

\section{Ricci Gauging}

Since the Weyl-potential $W_\mu$ is a vector and the Ricci curvature
is a second rank tensor it is clear that we can only find a
relation between them if we can construct a second rank tensor
from $W_\mu$. There are two local second rank tensors that
we can construct, namely $W_\mu W_\nu$ and $\nabla_\mu W_\nu$.
Let us now see how they transform under finite Weyl scaling. From 
(\ref{4.5}) we have
\beq
\Delta (W_\mu W_\nu)
=\om_\mu\om_\nu-(W_\mu\om_\nu+W_\nu\om_\mu),\mtx{where} \om_\mu=
\pa_\mu \om \label{5.1}
\eeq
has been introduced as not to clutter
the notation. To compute the finite Weyl-variation of
$\Delta (\nabla_\mu W_\nu)$ we need the variation of
$\Gamma_{\mu\nu}^\tau$ \cite{bide}
\beq
\Delta \Gamma_{\mu\nu}^\tau=g^{\tau\sigma}
\bigl(g_{\mu\sigma}\om_\nu+
g_{\nu\sigma}\om_\mu-
g_{\mu\nu}\om_\sigma\bigr),  \label{5.2}
\eeq             
which leads to
\beq
\Delta (\nabla_\mu W_\nu)=
\nabla_\mu\om_\nu-g_{\mu\nu}\;\om\!\cdot\!  \om
+2\;\om_\mu\om_\nu
+g_{\mu\nu}\;W\!\cdot\!  \om-(W_\mu\om_\nu+W_\nu\om_\mu).
\label{5.3}
\eeq
Each of the above variations \refs{5.1} and \refs{5.3} depend on $W_\mu$.
However, there is a combination
whose variation is independent of $W_\mu$. For this 
we first note that contracting (\ref{5.1}) we obtain
\beq
\Delta (g_{\mu\nu}W\!\cdot\!  W)=g_{\mu\nu}(\om\!\cdot\! 
\om-2W\!\cdot\!  \om).
\label{5.4}
\eeq
It is then easy to see that the variation of the tensor
\beq
\Om_{\mu\nu}[W]=\nabla_\mu W_\nu-W_\mu W_\nu+
{1 \over 2}g_{\mu\nu}\;W\!\cdot\!  W  \label{5.5}
\eeq    
is independent of $W_\mu$ and symmetric. More precisely,
\beq
\Delta\Om_{\mu\nu}[W]=-\bigl(\nabla_\mu\om_\nu-
\om_\mu\om_\nu+{1 \over 2}g_{\mu\nu}\;\om\!\cdot\!  \om\bigr)
=-\Om_{\mu\nu}[\om]. 
\label{5.6}
\eeq
Note the appearance of the same operator $\Om_{\mu\nu}$
on both sides of the equation. For $n\es 2$ we have 
\beq
\nabla^\mu W_\mu=-\nabla^2\omega=-\Omega^\mu_{\;\;\mu}
[\omega].\label{562}
\eeq

Because $\Om_{\mu\nu}[\om]$ depends only on the scale parameter
$\om$ it must have a geometrical significance. To see what geometrical 
object it corresponds to we recall the scale variation of the 
Ricci tensor \cite{bide}
\beq
\Delta R_{\mu\nu}= R_{\mu\nu}[e^{2\omega} g_{\mu\nu}]-R_{\mu\nu}[g_{\mu\nu}] 
= g_{\mu\nu}\nabla^2 \om
+(n-2)\Bigl(\nabla_\mu \om_\nu-\om_\mu \om_\nu+g_{\mu\nu}
\;\om\!\cdot\!  \om\Bigr).  \label{5.7}
\eeq
From \refs{5.7} we see that the tensor
\beq
S_{\mu\nu}=R_{\mu\nu}-{1\over 2(n-1)}g_{\mu\nu}R
\label{5.8}
\eeq
transforms under Weyl-scalings in the same way as $\Om_{\mu\nu}$, ie.
\beq
\Delta S_{\mu\nu}=(n-2)\Om_{\mu\nu}[\om].  \label{5.9}
\eeq
This shows for $n\neq 2$ that $\Om_{\mu\nu}[\om]$ is
proportional to the variation of $S_{\mu\nu}$. 
For $n\es2$ we have from (\ref{5.7})
\beq
R[e^{2\om}g_{\mu\nu}]-e^{-2\om}R[g_{\mu\nu}]=
2\nabla^2 \om=-2\Om^\mu_{\;\;\mu}[\om].  \label{5.10}
\eeq
We now can see when and why Weyl-gauging can
be replaced by a non-minimal coupling to the curvature: 
Weyl-gauging was introduced to compensate
the inhomogeneous part of the Weyl-variation of the kinetic
part of the action. But since, according to (\ref{5.8})
and (\ref{5.9}), the
Weyl-variations of $\Om_{\mu\nu}[W]$ and $\Om^\mu_{\;\;\mu}[W]$ are 
proportional to the variations of $S_{\mu\nu}$ and $R$
we see that {\it whenever} $W_\mu$ {\it appears in the
action only in the combination} $\Om_{\mu\nu}[W]$ {\it or} 
$\Omega^\mu_{\;\;\mu}[W]$ 
then it can be replaced by $S_{\mu\nu}$ or $R$.
From now on we refer to the compensation
of the inhomogeneous part of the kinetic terms by
$S_{\mu\nu}$ as {\it Ricci gauging} and we say that an 
action for which this is possible is {\it Ricci gaugeable}. \pan
What remains to be understood is under what conditions 
an action can be converted to a form in which
$W_\mu$ appears only in the combination $\Om_{\mu\nu}$.

\section{The Condition for Ricci Gauging}

It turns out that the necessary properties of the theory 
to be Ricci gaugeable are closely connected to conformal 
invariance. We therefore elaborate first 
on the connection between local Weyl-scalings and the 
conformal transformations, before obtaining the criterion for Ricci
gauging in the second part of this section. 
\vskip 0.3truecm \noindent
{\it Connection with Conformal Invariance} :\par\noindent
Within the
diffeomorphic context the equation defining the conformal
transformations
$x^\mu\ar y^\mu(x)$ is
\beq
{\pa x^\mu \over \pa y^\al}
{\pa x^\nu \over \pa y^\be}g_{\mu\nu}(x)
=\hat g_{\al\beta}(y) \quad \hbox{and} \quad
\hat g_{\al\be}(x)=e^{\hat \om(x)}g_{\al\be}(x). \label{6a.1}
\eeq
The $\hat\om$ in \refs{6a.1} form a subgroup of the group of local
Weyl-transformations 
that is induced by conformal transformations. We will call it
the {\it conformal Weyl-group}. We can characterize $\hat \omega$ using 
that we have from 
(\ref{6a.1})
\beq 
{\pa x^\mu \over \pa y^\al}{\pa x^\nu \over \pa y^\be}R_{\mu\nu}(x)=
\hat R_{\al\be}(y)=
R_{\al\be}[e^{2\hat\om}g_{\mu\nu}](y), \label{6a.2}
\eeq
which leads, in virtue of \refs{5.7} and \refs{5.9}, to the 
differential equation
\beq
(n-2)\Om_{\al\be}[\hat\om]=\hat S_{\mu\nu}-S_{\mu\nu},
\label{6a.3}
\eeq
for $n\neq2$ and
\beq
\nabla^2\hat\om=0,\label{dc2}
\eeq
for $n\es 2$. Next, we make a few comments on the solutions. 
The two-dimensional condition, which 
is a consequence of the conformal Killing equation, is solved by all 
harmonic functions. In higher dimensions the existence of global solutions 
of \refs{6a.3} is a non-trivial problem in general \cite{Kob}. 
In the flat-space 
limit $S_{\al\be}$ vanishes and the condition
(\ref{6a.3}) reduces to
\beq
\nabla_\mu\hat\om_\nu-
\hat\om_\mu\hat\om_\nu
+{1 \over 2}g_{\mu\nu}\;\hat\om\!\cdot\!\hat\om=0. \label{6a.4}
\eeq
In Cartesian coordinates the general solution of \refs{6a.4} is 
\beq
\hat\om(x)=\log(1-2c\!\cdot\!x+c^2 x^2),\quad n>2, \label{6a.5}
\eeq
where $c$ is an arbitrary constant vector. 

\vskip 0.3truecm \noindent
{\it The Criterion for Ricci gauging} :\par\noindent
After this digression we now obtain the criterion for 
Ricci gauging. Suppose that the Weyl-gauged action
${\cal A}$ permits Ricci gauging. We then have for the gauged action 
(with obvious modifications for $n\es2$)
\beq
{\cal A}(\phi, S_{\mu\nu})={\cal A}(\phi^\om,
S_{\mu\nu}+\Omega_{\mu\nu}[\om]),\label{ca}
\eeq
where $\phi^\om$ denotes the transformed fields. From this it follows 
in particular that if $\om$ fulfills $\Omega_{\mu\nu}[\om]\es0$ then 
the action is invariant even without gauging. But $\Omega_{\mu\nu}[\om]\es0$ 
is just the condition satisfied by the conformal Weyl-group \refs{6a.3} 
in flat space. Thus we have the result:
{\it A necessary condition for a Weyl-gauged action to admit
Ricci gauging is that it be conformally invariant in the flat-space limit}.

We now show that for actions which contain
only first derivatives of the conformally variant fields this condition is
also sufficient. Suppose an ungauged action $A_0$ is conformal 
invariant in flat space. For infinitesimal conformal
transformations we then have in Cartesian coordinates
\beq
\Delta A_0=\int\hat\om_{\mu}j^{\mu}
=c_{\mu}\int j^{\mu}=0,  \label{6.2}
\eeq
where $j^\mu$ is the virial current 
\beq
j^{\mu}=\pi_\nu\La^{\mu\nu}\phi;\qquad\pi^{\mu}
={\delta A\over \delta (\pa_\mu \phi)}  \label{6.3}
\eeq
introduced 
in section $3$ and $\hat\om=c_\mu x^\mu$ solves the linearized equation
(\ref{6a.4}). Because invariance is required off-shell,
ie. for arbitrary field configurations, (\ref{6.2})
implies that $j^\mu$ is a divergence
\beq
j^{\mu}=\pa_\nu J^{\mu\nu},  \label{6.4}
\eeq
where $J^{\mu\nu}$ is a tensor local in the fields, which 
we call the virial tensor.

For $n=2$ the condition is stronger because
the conformal Weyl-group contains
all harmonic functions.  Using this larger group we have
shown in the appendix that for $n=2$
\beq
J^{\mu\nu}=g^{\mu\nu} J \quad \hbox{and hence} \quad
j^\mu=\pa^\mu J , \label{6.5}
\eeq
where $J$ is a scalar. In other words for $n=2$
the virial current must not
only be a divergence but a gradient.

So far our results match those of Refs. \cite{coja,P}. However, we
can go farther because eqns. (\ref{6.4}) and (\ref{6.5})
are off-shell identities. As the actions considered contain
only first derivatives of the conformally variant fields the same is true for
$j^\mu$. Therefore if $J^{\mu\nu}$ contained any
derivatives of these fields, then (\ref{6.4}) and (\ref{6.5})
would imply a relation between the second derivatives of
the fields and lower ones, which is impossible for arbitrary field
configurations. Hence, we conclude that $J^{\mu\nu}$ depends only on
the conformally variant fields themselves and not on their derivatives.
Now we reverse the argument: since (\ref{6.4}) and (\ref{6.5})
involve only one derivative, they imply that
$j^\mu$ is at most linear in the first derivatives of the 
conformally variant fields.
{\it Thus, conformal invariance is possible only for actions which
are at most quadratic in the derivatives of the conformally variant fields}. 
Furthermore, in the case where the action is linear in the first
derivatives the same argument implies that the virial current
be zero. This situation occurs for the Dirac action.
Note that the dependence on the conformally invariant fields is not restricted 
by this argument. Examples where the Lagrangian is of higher order 
in the derivatives of the invariant fields are given by the Skyrme- and 
WZW-models.  

Because conformal invariance requires the action to be
at most quadratic in the first derivatives of the fields its variation under 
finite conformal transformations is then simply
\beq
\Delta A_0=\int\Bigl(\hat\om_\mu j^\mu +
\hat\om_\mu \hat\om_\nu T^{\mu\nu}\Bigr),  \label{6.6}
\eeq
where $T^{\mu\nu}$ does not contain any derivatives of the fields.
Using furthermore that $j^\mu$ is a total divergence we
have after a partial integration
\beq
\Delta A_0=\int \Bigl( -J^{\mu\nu}\pa_\mu \hat \om_\nu
+\hat\om_\mu\hat\om_\nu T^{\mu\nu}\Bigr).  \label{6.7}
\eeq
For $n\not=2$ we may use (\ref{6a.4}) to recast (\ref{6.7}) in the form
\beq
\Delta A_0=\int\,\hat\om_\mu\hat\om_\nu\Bigl(
T^{\mu\nu}-J^{\mu\nu}+{1 \over 2}g^{\mu\nu}J\Bigr),  \label{6.8}
\eeq
where $J$ denotes the trace of $J^{\mu\nu}$. Although $\hat\omega$ is not a 
completely arbitrary function, it depends on an arbitrary four-vector $c_\mu$. 
This allows us to  conclude from the vanishing of the variation for 
all field configurations that 
the integrand in \refs{6.8} must be zero (see appendix). Thus,
\beq
T^{\mu\nu}=J^{\mu\nu}-{1 \over 2}g^{\mu\nu}J 
\qquad \hbox{or} \qquad J^{\mu\nu}=T^{\mu\nu}-{1 \over n-2}g^{\mu\nu}T.
\label{6.9}
\eeq
Hence, invariance under finite transformations implies that the 
virial tensor $J^{\mu\nu}$ be a specific linear function of the
tensor $T^{\mu\nu}$ which appears in the quadratic expansion.
This puts strong algebraic conditions on the action as we
shall see later.\pan 
For $n=2$ we show in the appendix that invariance implies 
$T^{\mu\nu}=Kg^{\mu\nu}$, where the field independent
constant $K$ is uniquely determined by the virial current. We discuss this 
central term in section $6$. Here we take $K\es 0$. 

To continue, we return to the covariant formulation of the theory. The above 
analysis implies that the (minimal) covariant form of the action is 
at most quadratic in the derivatives of the conformally variant fields and 
that
\beq
j^\mu=\nabla_\nu J^{\mu\nu} \mtx{with} J^{\mu\nu}=T^{\mu\nu}-{1 \over 
n-2}g^{\mu\nu}T.   \label{8.9}
\eeq
Accordingly, if we Weyl-gauge the action we obtain
\beq
{\cal A}={\cal A}_0+\int \sqrt{g} \Bigl(W_\mu j^\mu +
W_\mu W_\nu T^{\mu\nu}\Bigr)  \label{6.10}
\eeq
and hence by using the same manipulations as were used
above we find that
\beq
{\cal A}={\cal A}_0+\int\sg \Bigl(-J^{\mu\nu}\,\nabla_\mu W_\nu
+W_\mu W_\nu\,T^{\mu\nu}\Bigr),  \label{6.11}
\eeq
which, using (\ref{6.9}), can be written as
\beq
{\cal A}={\cal A}_0+\int \sqrt{g}
J^{\mu\nu}\Om_{\mu\nu}[W].  \label{6.12}
\eeq
For $n=2$ the expression in (\ref{6.10}) then reduces to
\beq
{\cal A}={\cal A}_0-\int \sqrt{g}J\,\nabla\!\cdot\! W
={\cal A}_0-\int \sqrt{g} J\Om^\mu_{\;\;\mu}[W].  \label{6.13}
\eeq
Thus, for theories which are conformally invariant in the
flat-space limit the Weyl-potential only appears in the form
$\Om_{\mu\nu}[W]$ which is just the condition for Ricci gauging. 
Therefore we have proved the following: \pan
{\it A necessary and sufficient
condition for a scale-invariant action ${\cal A}$ to allow
for Ricci gauging is that the flat-space limit of the ungauged
action $A_0$ is conformally invariant. Furthermore the Ricci-gauging 
is achieved by} \refs{6.12} {\it respectively} \refs{6.13}.
\pan The improved energy-momentum tensor is then derived
from \refs{6.12} and \refs{6.13} respectively in the usual
way by a metric variation.

\section{Gravitational Anomalies in Two Dimensions}

We have seen in the last section that for $n\es 2$ the tensor $T^{\mu\nu}$ 
is of the form $T^{\mu\nu}\es Kg^{\mu\nu}$, with $K$ fixed by the group 
property of the conformal Weyl-transformations. We now have a closer look at 
this particular contribution to the conformal variation of the action. Since 
the linear term vanishes separately for $n\es2$ we have
\beq
\Delta_{\hat\om}{\cal{A}}=K\int\sg\;
\hat\om_\mu\hat\om^\mu, \label{ga1}
\eeq
with positive definite integrand, so that $\Delta_{\hat\om}{\cal{A}}$ 
is non-zero unless $\hat\om$ is constant. We then conclude that, for $K\neq 0$,
the action is conformally invariant only up to central 
(ie. field independent) terms. On the other hand, for arbitrary 
Weyl-transformations 
$g\ar e^{2\om} g$, 
falling off at infinity, the variation becomes after a partial integration
\beq
\Delta_\om{\cal{A}}=K\int\sg\;\om R-
K\int\om\nabla^2\om. \label{ga3}
\eeq
Hence, if we Ricci-gauge the action according to \refs{6.13}, 
then the resulting action is still Weyl-invariant up to a field independent 
contribution \refs{ga3}. The reader can easily convince
himself that taking a derivative of \refs{ga3} with 
respect to the metric 
leads to the trace formula \refs{1.2} 
which in turn determines the central charge of the Virasoro 
algebra \cite{Aff}. Now we see the connection 
between the restricted conformal invariance \refs{ga1}, the 
centre of the Virasoro algebra and the breaking \refs{ga3} 
of Weyl-invariance of partially Ricci-gauged conformal theories. 
The Liouville 
theory described in section $2$ nicely illustrates this connection. 
The ungauged 
model, obtained from \refs{2.1} by removing the coupling to the curvature 
scalar, 
is totally invariant under rigid scale transformations, but is invariant 
under arbitrary {\it conformal} Weyl-transformations
only up to the central term 
$\ha\int\hat\om_\mu\hat\om^\mu$. Discarding this term in the 
Weyl-gauging then leads to the partially Ricci-gauged action \refs{2.1} and 
the corresponding improved energy-momentum tensor satisfies the trace 
formula \refs{1.2}. So much for the classical theory. 
At the quantum level there is an additional central term due to 
the scale anomaly, which can be avoided only at the expense of 
reducing the diffeomorphism invariance \cite{Duff}. 
Although our approach so far was purely classical it naturally 
extends to the quantum theory replacing 
the currents by their expectation values. In particular, the operator
analog of \refs{6.6} 
\beq
\Delta {\cal W}=\int\,\hat\om_\mu\<j^\mu\>+
\hat\om_\mu\hat\om_\nu\<T^{\mu\nu}\>,  
\label{ww}
\eeq
is obtained by varying the Schwinger functional ${\cal W}$. 
It takes into account the scale anomaly of the measure which contributes 
a central term to \refs{ww}.

\section{Conformal Condition for Fields of Arbitrary Spin}

\vskip 0.3truecm \noindent 
{\bf Scalar Fields}
\vskip 0.2truecm \noindent
The standard rigid scale-invariant action for one scalar field in
$n\geq 3$ dimensions takes the form    
\beq
A=\int \left({1 \over 2}
\pa^a\phi\,\pa_a\phi-f 
\phi^{2n \over (n-2)} \right), \label{7.1}
\eeq
where the scale-dimension of the field is $d_\phi=(2-n)/2$ and
$f$ is a coupling constant. It is easy to see that the virial
current in this case is $j_\mu=d_\phi\pa_\mu \phi^2$. Thus
the action  is conformally invariant, and if we Weyl-gauge
it we obtain
\beqq
{\cal A}&=&\int \sg \left({1 \over 2}g^{\mu\nu}
(\pa_{\mu}+{d_\phi}W_\mu)\phi\;
(\pa_{\nu}+{d_\phi}W_\nu)\phi-f
\phi^{2n \over (n-2)}\right)  \nonumber \\
&=&\int \sg \left({1 \over 2}\pa^\mu\phi\,\pa_\mu\phi-f
\phi^{2n \over (n-2)}-{{d_\phi}\over 2}
\Omega^\mu_{\;\;\mu}\phi^2\right),  \label{7.2}
\eeqq
where $\Omega_{\mu\nu}$ is as in \refs{5.5} and can therefore be replaced 
by a multiple of $R$.

For $n=2$ scalar fields are dimensionless and no
invariant polynomial potential can be constructed.
However, definite scale dimensions can be assigned to
the exponentials of scalar fields and these exponentials
can be used to construct invariant potentials. For example the action 
\beq
A=\int \left({1 \over 2}\eta^{ab}\Bigl[\pa_a\theta\;
\pa_b\theta+ h_{\al\beta}(\hat
\phi)\pa_a\hat\phi^\al\;\pa_b\hat\phi^\beta\Bigr]-e^{\theta}
V(\hat \phi)\right), \label{7.3}
\eeq
where the fields $\hat \phi$ are conformal
scalars, is scale-invariant provided $e^\theta$
has scale dimension $d_{e^\theta}=-2$. 
The action (\ref{7.3}) is just the Liouville
action modified by the addition of some
scalar fields $\hat \phi$ whose potential is arbitrary. In this
case the virial current is $j_\mu=\pa_\mu \theta$ and hence
the full action  is conformally 
invariant. If we Weyl-gauge it we
obtain 
\beqq
{\cal A}\!\!\!&=&\!\!\!\int \sg \left({1 \over 2}g^{\mu\nu}
\Bigl[(\pa_\mu\theta-2W_\mu)(\pa_\nu\theta-2W_\nu)
+ h_{\al\beta}(\hat\phi)\pa_{\mu}\hat\phi^\al
\pa_{\nu}\hat\phi^\beta\Bigr]-e^{\theta}
V(\hat \phi)\right) \nonumber \\
\!\!\!&=&\!\!\!\int \sg \left({1 \over 2}g^{\mu\nu}
\Bigl[\pa_\mu\theta\;\pa_\nu\theta
+ h_{\al\beta}(\hat
\phi)\pa_{\mu}\hat\phi^\al\pa_{\nu}
\hat\phi^\beta\Bigr]-e^{\theta}
V(\hat \phi)-2\nabla\!\cdot\! W \theta
+2W\!\cdot\!  W\right). \label{7.4}
\eeqq
This is of the form of (\ref{6.13}) given in section 5.
The last term in \refs{7.4} is an example of a central term discussed 
in the last section. Dropping it the (partial) Ricci gauging then 
amounts to replacing the term
$\nabla\!\cdot\! W\theta $ by $R\theta$.

\vskip 0.4truecm \noindent
{\it Counter Example} :\vskip 0.2truecm \noindent
It is interesting to see what happens if we
allow the kinetic term of the Liouville field
to be multiplied by scalar fields. Consider for example
\beq
A=\int \left({1 \over 2}
\Bigl[\eta^{ab} h(\hat\phi)
\pa_a\theta\;\pa_b\theta +
h_{\al\be}(\hat\phi)\pa_a\hat\phi^\al
\pa_b\hat\phi^\be\Bigr]-e^{\theta}
V(\hat \phi)\right),  \label{7.5}
\eeq
where as before the $\phi$-fields are conformal scalars and
$e^\theta$ has scale dimension $d_{e^\theta}=-2$.
In this case the virial current is
$j_\mu=h(\hat \phi)\pa_\mu \theta$ and thus is {\it not} a 
total derivative. It follows that although the action is rigid
scale-invariant it is not conformally 
invariant, and if it is Weyl-gauged the Weyl-gauging cannot be
replaced by  Ricci gauging. Thus, even for actions
which are quadratic in the derivatives of scalar fields
rigid scale-invariance does not necessarily  imply
conformal invariance.\par

The case of spin ${1 \over 2}$ and spin $1$ fields has been widely treated 
in the literature, starting with \cite{coja}. Below we consider fields of any spin in $n$-dimensions. \vfill\break
\vskip 0.3truecm \noindent 
{\bf Fermions}
\vskip 0.2truecm \noindent
The Dirac Lagrangian for spin ${1 \over 2}$ fermions 
is found to be not only conformally invariant but locally Weyl 
invariant, i.e. has a virial current which is identically zero. 
We wish to show that the same is true for the Rarita-Schwinger 
Lagrangian for fermions of arbitrary spin. The Rarita-Schwinger fields 
describing fermions of spin $s+{1 \over 2}$ are of the form 
$\psi_{\alpha ij...k}(x)$, where $\alpha$ is a Dirac index 
and the $i,j..k=1...s$ are vector indices 
with respect to which $\psi$ is completely symmetric and 
traceless. The fields also satisfy the condition 
\beq
\gamma^i\psi_{ij...k}(x)=0  \label{101}
\eeq
Condition \refs{101} is not a mass-shell condition but an irreducibility 
condition that ensures that the spin of the field is exactly $s+{1\over 2}$, and not a mixture of this and lower spins.  
The Rarita-Schwinger Lagrangian for these fields is 
\beq
L(\psi)=\bar \psi^{ij...k}\gamma^\mu\nabla_\mu \psi_{ij...k}    
\label{102}
\eeq
and is the natural generalization of the Dirac Lagrangian. The 
scale-dimension is evidently $d_\psi=(1-n)/2$ and the virial current is 
\beq
j_\mu=\bar \psi^{ij...k}\Bigl(\gamma^\nu 
(d_\psi g_{\mu\nu}-\sigma_{\mu\nu})\Bigr)\psi_{ij..k}
-\sum \bar \psi^{i..j..k}\Bigl(\gamma^\nu(\tau_{\mu\nu})_j^{\;\;j'}\Bigr)
\psi_{i..j'..k}  \label{103}
\eeq
where the sum is over all the vector indices in turn and 
\beq
\sigma_{\mu\nu}={1 \over 4}[\gamma_\mu,\gamma_\nu] \qquad 
(\tau_{\mu\nu})_{jj'}=g_{j\mu}g_{\nu j'}-g_{j\nu}g_{\mu j'}. 
\label{104}
\eeq
The first term in \refs{104} is the term that would be obtained in spin ${1 \over 
2}$ case and vanishes in the usual manner because
\beq
\gamma^\nu\sigma_{\mu\nu}={(1-n)\over 2}\gamma_\nu =d_{\psi}\gamma_\nu 
\label{105}
\eeq
The interesting point, however, is that the other spin terms also 
vanish because 
\beq
\gamma^{\mu}(g_{j\mu}g_{\nu j'}-g_{j\nu}g_{\mu j'})=
\gamma_jg_{\nu j'}-\gamma_{j'}g_{j\nu}  \label{106}
\eeq
and these matrices annihilate the fields on account of the 
irreducibility condition \refs{101}. Thus, like the Dirac Lagrangian, 
the Rarita-Schwinger Lagrangian \refs{102} 
is not only conformally invariant but local-Weyl invariant. 
\vskip 0.3truecm \noindent
{\bf Bosons} 
\vskip 0.3truecm\noindent
The most general Lagrangian that is quadratic in the 
derivatives of the fields is 
\beq
L(\phi)={a \over 2}\nabla^p \phi^{ij...k}\nabla_p  \phi_{ij...k}
+ {b \over 2}\nabla^r \phi^{sj...k}\nabla_s \phi_{rj...k}
+ {c \over 2}\nabla_p \phi^{pj...k}\nabla^t \phi_{tj...k},\label{107}
\eeq
where $\phi_{ij...k}$ is a totally symmetric, traceless tensor and we discard 
possible potential terms which play only a passive role for the conformal conditions. 
The scale-dimension is evidently $d_\phi=(2-n)/2$. The case 
$a+b=c=0$ is the generalization of the Maxwell Lagrangian to 
arbitrary spin. Note that the last two terms in \refs{107} differ 
only by a total divergence. The virial current is by \refs{6.3}
\beq
j_\nu=d_\phi\pi_\nu^{ij...k}\phi_{ij...k}-
\sum \pi_\mu^{ij...k}(\tau_\nu^{\;\;\;\mu})_j^{\;\;j'}\phi_{i..j'..k}  
\label{109}
\eeq
where $(\tau_{\nu\mu})_{jj'}$ is given in \refs{104} and 
\beq
\pi_\nu^{ij...k}=a\nabla_\nu \phi^{ij...k}+b\nabla^i \phi_\nu^{\;\;j...k}
+c\nabla_p \phi^{pj...k}g_\nu^{\;\;i}  \label{108}
\eeq
is the canonical momentum. Substitution into \refs{107} yields 
after rearranging of the terms 
\beqq
j_{\nu}&=&{1 \over 2}(d_\phi a-b)\nabla_\nu (\phi^{ij...k} \phi_{ij...k}) 
+\Bigl[b(d_\phi +1-s)-sa\Bigr]\nabla_i(\phi^{ij...k}\phi_{\nu j...k})\nonumber\\&&+\Bigl[2sa+b(s-d_\phi)+c(d_\phi+n+s-2)\Bigr]\nabla_p \phi^{pj...k}\phi_{\nu j...k} \label{112}
\eeqq
Since the first two terms in \refs{112} are divergences and the last term 
is not it follows that the condition for conformal invariance is
\beq
2sa+b(s-d_\phi)+c(d_\phi+n+s-2)=0  \label{113}
\eeq
in which case the virial current is of the form \refs{6.4} with $J^{\mu\nu}$ given by 
\beq
J^{\mu\nu}={1 \over 2}(d_\phi a-b)g^{\mu\nu}\;\phi^{ij...k} \phi_{ij...k} 
+\Bigl[b(d_\phi+1-s)-sa\Bigr]\phi^{ij...k}\phi^\nu_{\;\;j...k}. \label{114}
\eeq
Thus conformal invariance reduces the number of parameters from three 
to two and if the total divergence (ratio of $b$ to $c$) is fixed in 
advance it reduces them from two to one. 
Local Weyl invariance requires that the virial current 
vanishes and thus requires
\beq
d_\phi a-b=0 \mtx{and} b(d_\phi+1-s) -sa=0,   \label{115}
\eeq
which, for $a,b\neq0$ becomes
\beq
(s-d)(d+1)=0.\label{115a}
\eeq
Because $d\leq0$ and $s>0$, \refs{115a} then implies $d\es-1$ or, 
equivalently $n\es4$. Hence, in $4$-dimensions a Weyl invariant action 
exists for any integer spin while in any other dimension 
there is no Weyl-invariant Lagrangian of the type \refs{107}! 
In the general case the Weyl-gauged action is given by \refs{6.12} which includes coupling to 
the Ricci scalar as well as Ricci tensor. 

We have seen in the previous section that the conformal 
properties of the Lagrangian depend on total divergence terms, i.e. 
they depend on the parameters $b$ and $c$ separately and not on the 
combination $b+c$. This may be surprising, but can be understood by 
noting that conformal transformations 
depend explicitly on the coordinates and so conformal {\it variations} 
of local quantities may produce divergent terms. 
This property was used in \cite{coja} to construct a $4$-dimensional spin 
$1$ Lagrangian that 
was conformally invariant but not locally Weyl invariant, namely the 
Lagrangian \refs{107} with $n=4$, $s=1$.\par 
\vskip 0.6truecm \noindent
{\bf Quadratic Rarita-Schwinger Fields}
\vskip 0.6truecm \noindent
Finally we consider Lagrangians 
that are quadratic in Rarita-Schwinger fields. In analogy to \refs{107} the 
most general such Lagrangian is 
\beq
L(\psi)={a \over 2}\nabla_p \bar \psi^{i...k}
\nabla^p \psi_{i...k}
+{b \over 2}\nabla_p \bar \psi^{i...k}
\nabla^i \psi^p_{\;\;j...k} +{c \over 2} \nabla_p\bar 
\psi^{pj...k}\nabla^t\psi_{tj..k}, \label{116}
\eeq
where the last two terms on the right-hand side differ only by a 
total divergence. The scale-dimension is $d_\psi=(2-n)/2$. Due to the 
irreducibility condition \refs{101}, the generator $\sigma_{\nu\mu}$ in 
$\Lambda^{\nu\mu}$ 
can be absorbed by substituting the integer spin $s$ by 
$s+\ha$ in the 
corresponding formulas for 
bosons in the last section, except for the term
\beq
{a \over 2}\Bigl[\nabla^\mu \bar \psi^{i...k}\sigma_{\nu\mu}\psi_{i...k}
-\bar \psi^{i...k}\sigma_{\nu\mu}\nabla^\mu\psi_{i...k}\Bigr].\label{116a}
\eeq
Conformal invariance then requires 
\beq
a=0 \mtx{and} b(s+\ha-d_\psi)+c(d_\psi+n+s-\frac{3}{2})=0.\label{117}
\eeq
Furthermore 
\beq
J^{\nu\mu}=\frac{b}{2}(d_\psi+\ha-s)
\Bigl[\bar\psi^{\mu}_{\;\;j...k}\psi^{\nu j...k}
+\bar\psi^{\nu}_{\;\;j...k}\psi^{\mu j...k}\Bigr]-\frac{b}{2}
g^{\nu\mu}\bar 
\psi^{j...k}\psi_{j...k}.\label{118}
\eeq
To summarize, the conformal condition puts two constraints on 
the three parameters $a,b,c$ and in general, for a fixed total 
divergence there is no Lagrangian that is conformally 
invariant. Only if the total divergence is chosen so that 
\refs{117} holds do we get conformal invariance. Furthermore, there is 
{\it no}  non-zero value of the parameters for which the Lagrangian 
\refs{116} is locally Weyl invariant. The Weyl-gauging is then 
given by \refs{6.12} with $J^{\nu\mu}$ from \refs{118}.

\section{Conclusions}
Exploiting the equivalence 
between the conformal group and the conformal Weyl-group we have reformulated 
a previous criterion for conformal invariance in flat space and obtained 
a purely algebraic criterion which permits a systematic 
analysis. In particular, any conformally invariant action without 
higher derivatives in the conformal variant fields must be at most
quadratic in the first derivatives of variant fields. For such actions we have obtained the conditions for scale-invariance to 
imply conformal invariance for any spin in arbitrary dimensions. 

We have shown that the flat space conformal theories
are the only ones whose diffeomorphic versions can be made
Weyl-invariant (Weyl-gauged) by non-minimal coupling to the curvature. 
In $n$ dimensions the coupling is to the Ricci-tensor as well as to the 
Ricci-scalar. The Dirac action for half-integer spin fields is 
Weyl-invariant even without gauging due to the vanishing of the virial 
current. An interesting situation arises for integer spin $s\neq0$: 
For $n\es4$ a local Weyl-invariant action exists for arbitrary $s$, but for 
$n\neq4$ no such action exists for any $s$. 

In two dimensions the non-minimal coupling is related to the central term in 
the conformal variation of the Lagrangian 
and to the centre of the Virasoro algebra. Furthermore, the Weyl-gauged
action leads automatically to the improved energy-momentum
tensor upon variation with respect to the metric. Because the scale 
parameter is parity invariant, the improvement terms induced by Weyl-gauging 
are parity even. Parity violating improvements which are compatible with 
the conformal group in two dimensions are therefore not induced by 
Weyl-gauging. But they can be obtained by coupling each chiral component to 
gravity separately \cite{Jade}. 

\section*{Acknowledgments}
I. S. and C. W. have been partially supported by 
the Swiss National Science Foundation. L. O'R. would like to thank 
R. Jackiw for some illuminating comments and references. I. S. 
is indebted to D. Catalano for helpful comments on global conformal 
transformations.

\section*{Appendix}
\setcounter{equation}{0}

\vskip 0.4truecm \noindent {\it Supplementary Condition on the 
Virial Current in Two Dimensions} :\vskip 0.2truecm \noindent

Using the general divergence condition (\ref{8.9}) we see
that for $n=2$ the condition on the virial
current may be written as
\beq
\int j^\mu\hat\om_\mu =-\int J^{\mu\nu}
\nabla_\nu\hat\om_\mu =0,  \label{C.1}
\eeq
where $\hat\om$ is any harmonic function. Furthermore, by the
arguments given earlier, if we assume that the action depends only
on the fields and their first derivatives the virial tensor 
$J^{\mu\nu}$ cannot depend on
the derivatives. Now, choosing $\hat\om=\hat\om(x^+)$
we see that (\ref{C.1}) becomes
\beq
\int dx^+dx^-  J^{++}\nabla_+^2 \hat \om(x_+)=0
\label{C.2}
\eeq
which implies that
\beq
\int dx_-  J^{++}=0 \quad \hbox{or} \quad
J^{++}=\pa_- S^{++}  \label{C.3}
\eeq
for some local function $S$. But since $J^{++}$
does not depend on the derivatives of the fields
(\ref{C.3}) cannot hold for arbitrary field
configurations unless $J^{++}=0$.
Similarly $J^{--}=0$. It follows that
\beq
J^{\mu\nu}=g^{\mu\nu}J  \label{C.4}
\eeq
as required. Note that (\ref{C.1}) imposes no further
condition on $J$ because
$\hat \om$ is harmonic.

\vskip 0.4truecm \noindent {\it Vanishing of the Integrand in} \refs{6.8} :
\vskip 0.2truecm \noindent
%\section{}

Let $\hat\om$ be as in \refs{6a.5}. Then
\beq
\int \hat\om_\mu\hat\om_\nu S_{\mu\nu}d^nx=
\int {x_\mu x_\nu \over x^4}S_{\mu\nu}(x+e)d^nx \mtx{where} e_\mu
=\frac{c_\mu}{c^2}. \label{A.1}
\eeq
Multiplying \refs{A.1} by $e^{ike}$ and integrating over $e$ then 
leads to the identity 
\beq
S_{\mu\nu}(k)\int  {x_\mu x_\nu \over x^4}e^{-ikx}d^nx=0.\label {A.2}
\eeq
Performing the integration over $x$ leads to 
\beq
\frac{2}{n}k^2 S^\mu_{\;\;\mu}+(2-n)k^\mu k^\nu\hat S_{\mu\nu}=0,   \label{A.3}
\eeq
where $S^{\mu\nu}\es\frac{1}{n}S^\mu_{\;\;\mu}g^{\mu\nu}+\hat S^{\mu\nu}$.
Because \refs{A.3} is an off-shell identity
we conclude that $\hat S_{\mu\nu}\es 0$ and 
$S^\mu_{\;\;\mu}$ can at most be constant, 
which is excluded for $n>2$ on dimensional grounds.\par 

\vskip 0.4truecm \noindent {\it The case} $n\es 2$ :
\vskip 0.2truecm \noindent

For $n\es 2$ we obtain a relation between $j^\mu$ and $T^{\mu\nu}$ from 
\refs{6.6}. Because the linear term in \refs{6.6} vanishes separately, 
the tensor $T^{\mu\nu}$ can be at most constant ie. $T^{\mu\nu}\es 
Kg^{\mu\nu}$. Then, using 
the group property $\Delta_{\hat\om^1+\hat\om^2}\es\Delta_{\hat\om^2}
\Delta_{\hat\om^1}$ we obtain the relation
\beq
\int\;\hat\om^1_\mu(\Delta_{\hat\om^2}j)^\mu=
2K\int\;\hat\om^1_\mu\hat\om^{2\mu}, 
\label{n2t}
\eeq
which is the analog of \refs{A.1} for $n\es2$ and implies 
\beq
\frac{\de j^\mu(x)}{\de\hat\om_\mu(y)}=Kg^{\mu\nu}\de(x-y). \label{n3t}
\eeq
Note that \refs{n3t} requires the virial current $j^\mu$ to be linear 
in the conformally variant fields with constant coefficients. This property is 
not automatic, even for quadratic Lagrangians as is shown by the counter 
example in section $7$.

\end{document}